\documentclass[onecolumn, a4paper]{article}

\usepackage{authblk}
\usepackage{amsmath}
\usepackage{graphicx}
\usepackage{caption}
\usepackage{float}
\usepackage{epstopdf}

\usepackage{siunitx}
\DeclareSIUnit[number-unit-product = {}]
\uL{\micro\liter}
\DeclareSIUnit[number-unit-product = {}]
\umol{\micro \mole}
\DeclareSIUnit[number-unit-product = {}]
\uM{\micro M}

\usepackage{bpchem}   
\title{Automated tracking of colloidal clusters with sub-pixel accuracy and precision}

\author[1]{Casper van der Wel}
\author[1, *]{Daniela J. Kraft}

\affil[1]{Soft Matter Physics, Huygens-Kamerlingh Onnes Laboratory, Leiden University, PO Box 9504, 2300 RA Leiden, The Netherlands}
\affil[*]{To whom correspondence should be addressed. E-mail: kraft@physics.leidenuniv.nl}
\date{}

\begin{document}

\maketitle

\begin{abstract}
Quantitative tracking of features from video images is a basic technique employed in many areas of science.	
Here, we present a method for the tracking of features that partially overlap, in order to be able to track so-called colloidal molecules. Our approach implements two improvements into existing particle tracking algorithms. Firstly, we use the history of previously identified feature locations to successfully find their positions in consecutive frames. Secondly, we present a framework for non-linear least-squares fitting to summed radial model functions and analyze the accuracy (bias) and precision (random error) of the method on artificial data. We find that our tracking algorithm correctly identifies overlapping features with an accuracy below \SI{0.2}{\percent} of the feature radius and a precision of 0.1 to 0.01 pixels for a typical image of a colloidal cluster. Finally, we use our method to extract the three-dimensional diffusion tensor from the Brownian motion of colloidal dimers.
\end{abstract}

\textbf{This document is a preprint version: for the (revised) accepted version, please refer to the publisher's website via the following DOI: 10.1088/1361-648X/29/4/044001}

\section{Introduction}
Extracting quantitative information about the position and motion of features in video images is often key to understanding fundamental problems in science.
For example, the tracking of colloidal hard spheres in three-dimensional confocal images has provided important insights into phenomena such as melting, crystallization, and the glass transition \cite{Murray1996, Weeks2000, Kegel2000, Dinsmore2001, Meng2014}.
Biophysical experiments such as the investigation of cell mechanics by microrheology \cite{MacKintosh1999, Tseng2002} or the measurement of single biomolecule mechanics using optical or magnetic tweezers \cite{Neuman2008} rely on the precise positional measurement of single colloidal particles.
Moreover, the tracking of single proteins in live cells provided a powerful tool for understanding biological processes \cite{Gelles1988, Yildiz2003}, and eventually lead to the development of super-resolution microscopy techniques such as PALM and STORM \cite{Betzig2006, Huang2007}.
Crucial for these studies is a method to extract trajectories of features from video images, which has been described extensively in colloidal science \cite{Crocker1996, Savin2005} as well as in single molecule tracking \cite{Ghosh1994, Saxton1997a, Ober2004, Smith2010}.

Most single particle tracking algorithms have been designed for spherical features, as it is the most common type of signal.
Recent developments in colloidal synthesis \cite{Manoharan2003, Kraft2009, Meester2016} provide means to assemble spheres in so-called colloidal molecules.
Single particle tracking of these clusters of spheres will provide insights into the role of anisotropy in for instance crystallization and diffusion \cite{Glotzer2007, Hunter2011, Edmond2012a}.
As the basic building blocks of these studies contain closely spaced particles, a robust automated method is required to perform accurate particle tracking on partially overlapping features.

Automated methods for single-particle tracking follow roughly the following pattern: an image with features of interest is first preprocessed, then single features are identified in a process called ``segmentation'', these feature coordinates are refined to sub-pixel accuracy, and finally the features are linked to the features in the previous image. Iteration of this algorithm over a sequence of images results in particle trajectories that can be used for further analysis.
Although this method has proven itself as a robust and accurate method \cite{Cheezum2001, Jenkins2008}, issues arise when features become so closely spaced that their signals overlap.
This essentially limits studies to dilute systems, repelling particles, or model systems with very specific characteristics such as index-matched and core-shell fluorescent particles \cite{VanBlaaderen1995, Besseling2009}.

In particular, overlapping feature signals give rise to two complications:
firstly, the segmentation step regularly recognizes two closely spaced features as one feature due to the overlap of signals.
In order to identify the trajectories of closely spaced features completely, tedious frame-by-frame manual corrections are necessary, prohibiting the analysis of large data sets.
In super-resolution microscopy methods, reported approaches to solve this issue are repeated subtraction of point-spread functions of detected features \cite{Serge2008}, or advanced statistical models classifying merge and split events \cite{Jaqaman2008}.
Notably, these tracking methods do not use all the available information: as the feature locations are known in the previous frame, the segmentation of the image may be enhanced using the projected feature locations.
Here we will present a fast and simple method for image segmentation that makes use of this history of the feature locations. We will test this method on artificial images and experimental data of colloidal dimers.

A second issue that arises when two feature signals overlap is that their refined coordinates will underestimate the separation distance.
Especially the commonly employed center-of-mass centroiding suffers from this systematic ``overlap bias'', leading to on apparent attraction between colloidal particles \cite{Jenkins2008, Baumgartl2006}.
For fluorescence images, this issue can be addressed by least-squares fitting to a sum of Gaussians, which has been reported as a way to measure the distance between overlapping diffraction limited features \cite{Gordon2004, Qu2004a}.
Here, we will apply this method to images with features that are not diffraction limited.
We conduct systematic tests on the accuracy (bias) and precision (random error) of the obtained feature positions.

To demonstrate the new automated segmentation and refinement methods, we will apply it to three-dimensional confocal images of a diffusing colloidal cluster consisting of two spheres and use the obtained trajectories to extract its diffusion tensor.

\section{Method}

\subsection{Segmentation}
\begin{figure}[h]
	\centering
	\includegraphics{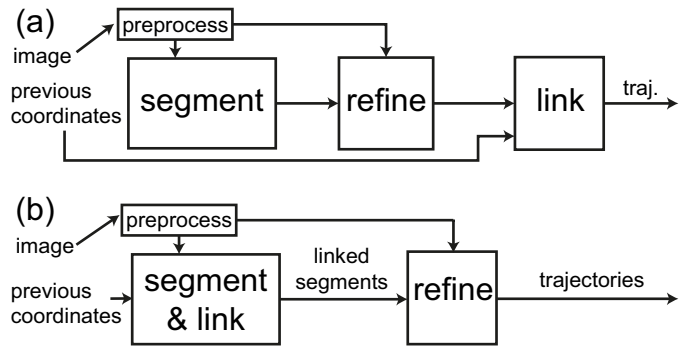}
	\caption{Schematic of the particle tracking of a single frame in (a) the Crocker-Grier algorithm and (b) the new algorithm. In the Crocker-Grier algorithm (a) the image is preprocessed and segmented. From the segments and the preprocessed image, a refinement step is done. Finally, subsequent coordinates are linked together with the coordinates in the previous frame. In our new algorithm (b) the image is preprocessed and segmented, making use of the knowledge of the previous coordinates. The linked segments are refined afterwards.}
	\label{fig:flowchart}
\end{figure}

As our algorithm for single particle tracking is based on the widely employed algorithm by Crocker and Grier \cite{Crocker1996}, we will first introduce their algorithm and call it ``CG-algorithm''. Throughout this work a Python implementation of this algorithm, Trackpy \cite{trackpy031}, was used for comparison. The CG-algorithm consists of four subsequent steps: preprocessing, feature segmentation, refinement, and linking. See Figure \ref{fig:flowchart}(a) for a schematic overview.

The preprocessing consists of noise reduction by convolution with a \SI{1}{px} sized Gaussian kernel and background reduction by subtracting a rolling average from the image with kernel size $2R + 1$. The length scale $R$ is chosen just larger than the feature radius. The subsequent segmentation step finds pixels that are above a given relative intensity threshold and are local maxima within a certain radius $S$. The length scale $S$ is the minimum allowed separation between particles. After the refinement step (see next section) the linking connects the features in frame $i$ with features in frame $i - 1$ by minimizing the total displacement between the frames. Between two frames, particles are allowed to move up to a maximum distance $L$.

\begin{figure}[h]
	\centering
	\includegraphics{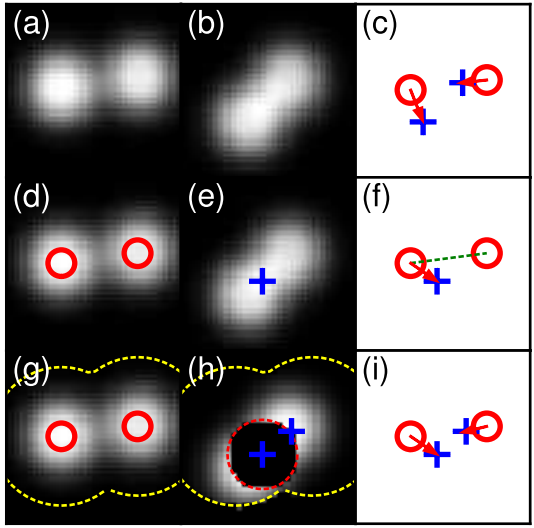}
	\caption{Artificial example to illustrate the integrated segmentation and linking step. In (a) and (b) two subsequent computer-generated frames are shown and in (c) the corresponding true feature locations, with the frame 1 features in red circles and the frame 2 features in blue crosses. Links are indicated by red arrows. In (d) frame 1 is shown again, overlaid with its feature coordinates in red circles and in (e) the result of the initial feature finding is indicated by a blue cross on top of frame 2. (f) The subnet is formed by the linking candidates. Additionally, the green dashed line denotes a distance between features that is less than $2L$. Therefore these features could belong to a single subnet via a missing feature. (g) Subsequently, a region is defined up to distance $L + S$ from the features in frame 1 (dashed yellow line), that is used to (h) mask frame 2. In this step, also all features that were found already are masked up to $S$, which enables the detection of the second feature that is less than distance $L$ from the features in frame 1 and farther than $S$ from other features in frame 2. The newly found feature is then added to the subnet so that the linking can be completed (i).}
	\label{fig:relocate}
\end{figure}

In this process, each frame is treated individually: only during the final step (linking), features are connected into trajectories. We rearranged this process such that the information about the particle locations in the previous frame is used already in the segmentation. This allows us to project the expected feature locations in consecutive frames and therefore increase the success rate of segmentation. See Figure \ref{fig:flowchart}(b) for a schematic overview. We describe the new segmentation algorithm here using a minimal example of two closely spaced features in two subsequent frames, which can be generalized to an arbitrary number of features in any number of frames. See Figures \ref{fig:relocate}(a)-(c).

We will assume that feature finding and refinement was performed successfully on the previous frame (Figure \ref{fig:relocate}(d)). The current frame is first subjected to gray dilation and thresholding step, just as in the CG-algorithm. Because features are closely spaced in the frame 2, this leads to segmentation into only one single feature (Figure \ref{fig:relocate}(e)).

Then a part of the linking step is executed: features are divided into so-called subnetworks. This is a necessary step in the CG algorithm to break the $\mathcal{O}(N!)$ sized combinatorial problem of linking two sets of $N$ features into smaller parts. First, linking candidates are identified using a kd-tree \cite{trackpy031, Maneewongvatana1999}. Linking candidates for features in frame 1 are features that are displaced up to a distance $L$ in frame 2 and vice versa. Then subnetworks are created such that all features that share linking candidates are in the same subnetwork. For a sufficiently large distance $L$, all features in Figure \ref{fig:relocate}(f) belong to the same subnet: the feature in frame 2 is a linking candidate for both features in frame 1.

From the subnetworks, the number and estimated location of missing features is obtained ``for free'': if a subnetwork contains fewer particles in frame 2 than in frame 1, there must be missing features in its vicinity. To account for the possibility that a missing feature could connect two subnetworks, we combine subnetworks if they are less than distance $2L$ apart in frame 1 whenever missing features are being located.

In order to estimate the location of the missing features, a region up to distance $L + S$ around the features in the previous frame is masked in the current frame (dashed yellow line in Figures \ref{fig:relocate}(g)-(h)). Subsequently, all already found features are masked up to a radius of $S$ (Figure \ref{fig:relocate}(f)). This enables us to find local maxima that are further than distance $S$ from all other features in the current frame and closer than distance $L$ from the features in the previous frame. From the masked subimage, local maxima are obtained again through gray dilation and thresholding. After this, feature selection filters can be inserted in order to select appropriate features, for example with a minimum amount of integrated intensity. Then the new feature is added to the subnetworks and linking is completed by minimizing the total feature displacement (Figure \ref{fig:relocate}(i)).

By performing the linking during the segmentation process, additional information is taken into account: not only the present image is used to identify the features, but also the coordinates from the previous frame. Therefore, we expect a higher number of correctly identified feature positions for the combined linking and segmentation method. Because all the computationally intensive tasks were already present in the original algorithm, the execution time of our new algorithm was observed to be similar. 

\subsection{Refinement}
Subpixel accuracy and precision is a key feature of single particle tracking. Although the size of a single pixel is diffraction limited to approximately \SI{200}{nm}, localization precisions down to \SI{1}{nm} have been reported \cite{Ober2004, Cheezum2001}. These subpixel feature locations are obtained by starting from an initial guess supplied by the segmentation step, which is then improved in the so-called ``refinement'' step. Here, we will describe a general-purpose framework for refinement of overlapping features using non-linear least squares fitting to summed radial model functions.

We will compare this method to the center-of-mass centroiding that is present in the CG algorithm \cite{Crocker1996}. For radially symmetric features, the feature position is given by its center-of-mass. Due to its simplicity and computational efficiency, this method is a preferred choice for many tracking applications. In the center-of-mass refinement, the center coordinate $\vec{c}$ of the feature is obtained iteratively from the image $I(\vec{x})$, such that:

\begin{equation}
\sum_{\mathrm{dist}(\vec{x}, \vec{c}) \leq R} I(\vec{x}) (\vec{x} - \vec{c}) = 0.
\label{eq:CoM}
\end{equation}

Non-linear least squares fitting to a model function is conceptually different, since it goes beyond assuming only feature symmetry and requires knowledge on the feature shape. If image noise is uncorrelated and normal distributed, this method gives the maximum likelihood estimate of the true centroid. Although this assumption is not strictly valid \cite{Savin2005, Ober2004}, the precision of this method is generally higher than the center-of-mass method when the image is subject to noise \cite{Cheezum2001}. By simultaneously fitting a sum of multiple model functions, this method can be extended to tracking multiple overlapping features \cite{Gordon2004, Qu2004a}.
We employ this approach here and formulate the feature model function $F$ in the following way:

\begin{align}	
F(\vec{x}, \vec{c}, A, \vec{\sigma}, \vec{p}) &=
\begin{cases}
A \cdot f(r(\vec{x}, \vec{c}, \vec{\sigma}), \vec{p})&\mathrm{dist}(\vec{x}, \vec{c}) \leq R\\
	0&\mathrm{otherwise}\label{eq:generic_model}
\end{cases},\\
r^2(\vec{x}, \vec{c}, \vec{\sigma}) &= \sum_{j = 1}^D \left( \frac{x_j - c_j}{\sigma_j} \right)^2.
\label{eq:r}
\end{align}

Here, $\vec{x}$ is the image coordinate, $\vec{c}$ the feature center, $A$ its intensity,  $\vec{\sigma}$ its radius, and $f$ a model function of a single feature, which is a function of $r$ and a list of parameters $\vec{p}$. The reduced radial coordinate $r$ is defined for any number of dimensions $D$ and allows for anisotropic pixel sizes through the vector nature of $\vec{\sigma}$. The feature model function is defined only up to distance $R$ from the feature center. It is in principle possible to use any function for $f$ and apply it to images with different signal intensities and physical pixel sizes through the separate parameters $A$ and $\vec{\sigma}$. In this article, we limit ourselves to the Gaussian function $f(r) = \exp{[-r^2]}$ so that we do not have extra parameters $\vec{p}$. We keep $\vec{\sigma}$ constant and allow $\vec{c}$ and $A$ to be optimized.

The model image is constructed by the summation of the individual features, which are each only defined within a region with radius $R$. This additivity is a good assumption for fluorescence microscopy techniques \cite{Jenkins2008}. We add a fixed background signal $B$, which we keep constant within each cluster of overlapping features, but we allow it to vary between clusters to account for spatially different background values. For an image or video consisting of $N$ features, the following ``objective function'' is minimized:

\begin{equation}
\sum_{\vec{x}}
\left( I(\vec{x}) - B - \sum_{i = 1}^N 
F (\vec{x}, \vec{c_i}, A_i, \vec{\sigma}_i, \vec{p}_i)\right)^2.
\label{eq:global_obj}
\end{equation}

The feature model function $F$ is defined by Eq. \ref{eq:generic_model}. If all features are separated by more than $2R$, this minimization can be separated into $N$ single feature problems. However, when features have overlapping regions, their objective functions cannot be separated and have to be minimized simultaneously. We separate the full image objective function (Eq. \ref{eq:global_obj}) into groups (``clusters'') using the kd-tree algorithm \cite{Maneewongvatana1999}.
Each of the resulting cluster objective function is minimized using the Sequential Linear Least Squares Programming (SLSQP) algorithm \cite{Kraft1988} interfaced through the open-source Python package SciPy \cite{scipy}.
This SLSQP algorithm allows for additional constraints and bounds on the parameters. We use bounds to suppress diverging solutions and constraints to for example fix the distance between two features to a known value.
The optimizer is supplied with an analytic Jacobian of Eq. \ref{eq:global_obj} to increase performance.

The here described framework of feature refinement in principle allows refinement of any feature that can be described by a radial function. Although less computational efficient than the conventional refinement by center-of-mass, it can take into account feature overlap and additionally allows for constraints on parameters.

\subsection{Testing methods}
The above described methods for single particle tracking were tested quantitatively on both artificial and experimental data. Artificial images were generated by evaluating the following analytical functions for disc- and ring-shaped features on an integer grid:

\begin{align}
f_{disc}(r, d) &= \begin{cases}
\exp{\left[-\left(\frac{r - d}{1 - d}\right)^2\right]}&r \ge d\\
1&\mathrm{otherwise}\\
\end{cases},
\label{eq:disc}\\
f_{ring}(r, t) &= \exp{\left[-\left(\frac{r - t - 1}{t}\right)^2\right]}.
\label{eq:ring}
\end{align}

\begin{figure*}
	\centering
	\includegraphics{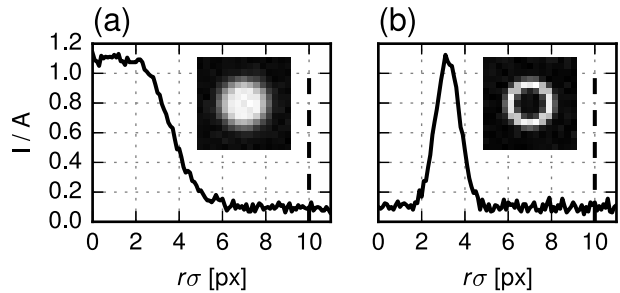}
	\caption{Radial functions of (a) a disc-shaped and (b) a ring-shaped model feature, generated with Equations \ref{eq:disc} and \ref{eq:ring} with parameters $\sigma = 4$, $d = 0.5$, and $t = 0.2$. The insets show the corresponding single-feature images. Poisson-distributed noise was added to each feature.}
	\label{fig:model_features}
\end{figure*}

Here, the reduced radial coordinate $r$ is given by Eq. \ref{eq:r}, $d$ is the solid disc radius in units of $\sigma$, and $t$ is the ring thickness in units of $\sigma$. The true feature location $\vec{c}$ was generated at a random subpixel location. Unless stated otherwise, we chose $d = 0.5$, $t = 0.2$, $\sigma = 4$, and $A = 160$. See Figure \ref{fig:model_features} for two example model features generated with these parameters.
Images were discretized to integer values and a Poisson distributed, signal-independent background noise with a mean intensity of $N = 16$ is added to each image. The signal-to-noise ratio is defined as $A / N$. 
Each refinement test was performed on 100 images having two overlapping features with a given center-to-center distance and random orientations. In order to ensure that the choice of initial coordinates did not affect the refined coordinate, we generated the initial coordinates randomly within \SI{2}{px} from the actual coordinate. 

Experimental measurements on colloidal particles were performed with an inverted Nikon TiE microscope equipped with a Nikon A1R resonant confocal scanhead scanning lines at \SI{15}{kHz}. For the two-dimensional diffusion measurements, we used a 20x objective (NA = 0.75), resulting in a physical pixel size of \SI{0.628}{\um}. For the three-dimensional measurements, a 100x (NA=1.45) oil immersion objective was used, resulting in an XY pixel size of \SI{0.166}{\um}. A calibrated MCL NanoDrive stage enabled fast Z stack acquisition with a Z step size of \SI{0.300}{\um}. As the objective immersion liquid ($N_D = 1.515$) is closely matched with the sample solvent ($N_D = 1.49$), this step size equals the physical pixel size in Z direction within an error of \SI{5}{\percent} \cite{Besseling2015}. We acquired 5.13 three-dimensional frames per second with a size of 512x64x35 pixels (x-y-z).

For two-dimensional diffusion measurements we employed samples consisting of partially clustered TPM (\IUPAC{3\-(tri\|methoxy\|silyl)\-propyl\|metha\|crylate}) colloids with a diameter of \SI{2.05}{\um} containing a FITC (fluorescein \IUPAC{iso\|thio\|cyanate}) fluorescent marker, as described in \cite{VanderWel2016a}. Particles were confined to the microscope coverslip through sedimentation.

The samples for three-dimensional measurements consisted of core-shell RITC (rhodamine B \IUPAC{iso\|thio\|cyanate}) labeled PMMA (\IUPAC{poly\|methyl\|methacrylate}) colloidal clusters that were synthesized via an emulsification-evaporation method according to \cite{Manoharan2003}. The average distance between the two constituent spheres of radius \SI{1.87+-0.06}{\um} in a cluster is \SI{1.58+-0.12}{\um}, determined by Scanning Electron Microscopy using an FEI NanoSEM at \SI{15}{kV}. The clusters were both index and density matched using a mixture of cyclohexyl bromide and cis-decalin in a weight ratio of 72:28 and imaged in a rectangular capillary, similar to experiments described in \cite{Kraft2013a}.

The Python code on which this work is based is available on-line\footnote{https://github.com/caspervdw/clustertracking} and will be integrated into a future version of Trackpy \cite{trackpy031}, that is available through Conda as well as through the Python Package Index. All tests described in this work are implemented as ``unittests'' that ensure the correct functioning of the code on each update.

\section{Results and Discussion}

\subsection{Segmentation and Linking}
\begin{figure*}
	\centering
	\includegraphics[width=\textwidth]{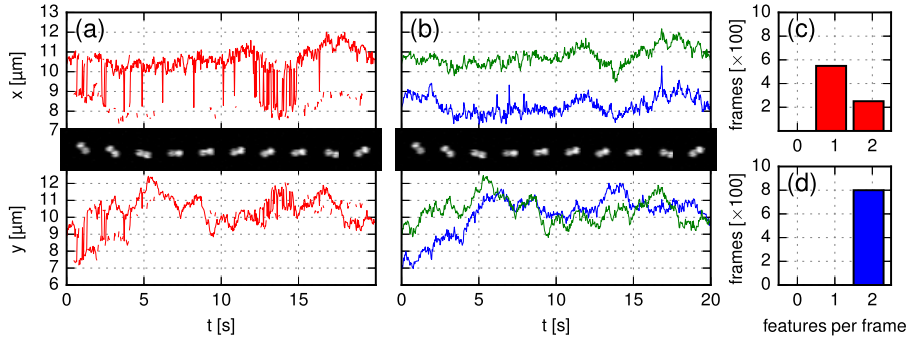}
	\caption{Segmentation of an experimental video image of a colloidal dimer. In (a) the x and y coordinates obtained using the CG algorithm are shown. The corresponding a histogram of features per frame is displayed in (c). As roughly two third of the frames had only one feature, trajectories cannot be identified. In (b) the trajectories were obtained using the integrated segmentation and linking algorithm. As all frames had two features (see the histogram in (d)), trajectories were identified completely. In these plots, coordinates were refined using least-squares fitting to a sum of Gaussians.}
	\label{fig:dimer}
\end{figure*}

As described in the method section, the integrated segmentation and linking step extends the frame-by-frame segmentation used in the CG algorithm in such a way that it makes use of the history of feature locations. 
In order to test the effect of our extension, we compared the segmentation in the CG algorithm with our integrated segmentation and linking on experimental video images. The video images contain a single diffusing colloidal dimer, which consists of two permanently connected spheres. The identified trajectories for 800 frames are displayed in Figure \ref{fig:dimer}. Clearly, by taking into account the history of the feature positions, the dimer positions can be tracked significantly better: for the new algorithm two features were detected in all of the 800 frames, while for the CG algorithm, only one third of the frames had 2 features, resulting in short disconnected trajectories that appear to hop between two feature locations.

The here described extension of segmentation increases the number of correctly segmented features significantly. It has to be noted though that the segmentation of the first frame is not enhanced by our method because of the lack of information on the previous feature positions. Generally, there is a start-up period of a few frames in which the number of correctly segmented features increases. These potentially incorrectly tracked frames can be ignored for most tracking applications. For cases where the first frames are relevant, the algorithm could be ran backwards from the first correctly segmented frame.

\subsection{Refinement}
After the segmentation step, the subpixel position is obtained in the refinement step. In this section we will analyze the effect of signal overlap on the accuracy and precision in the refined feature coordinates using both center-of-mass and the here described least-squares fitting to sums of model functions. We define the accuracy or bias as the mean difference between the measured and the true value. The precision is the random deviation around the measured average, which we calculate with the root of squared deviations from the measured average.

\begin{figure}
	\centering
	\includegraphics{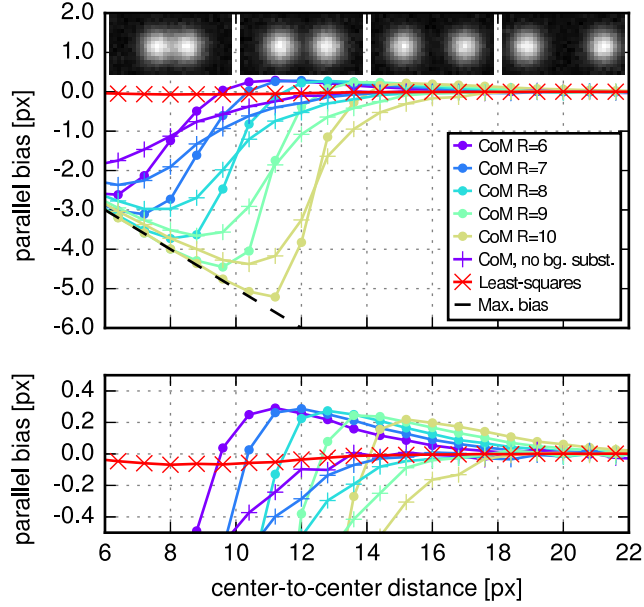}
	\caption{The effect of feature overlap on the bias in the parallel coordinate. The bias is negative when features appear too close together. In both graphs, the bias in the parallel coordinate as a function of the center-to-center distance is shown, for two Gaussian features with $\sigma = 4$ and signal-to-noise ratio $\mathrm{S/N} = 10$. The bias for the center-of-mass (CoM) refinement is shown for mask radius $R$ from 6 to 10, both with rolling average background subtraction (denoted with dots) and without (denoted with crosses). The bias for the least-squares fitting to a sum of Gaussians method is denoted with tilted crosses. The dashed black line denotes the bias at which features are detected precisely in between the two actual feature positions.}
	\label{fig:refine_distance}
\end{figure}

First, we took two Gaussian model features (Eq. \ref{eq:disc} with $d = 0$) and varied their spacing. See Figure \ref{fig:refine_distance}. The deviations of the obtained positions are measured parallel and perpendicular to the line connecting the two actual feature positions. We found for both refinement methods that there is no bias in the perpendicular coordinate. For the parallel coordinate, however, we found a clear difference between the two refinement methods: in center-of-mass centroiding, the parallel coordinate was negatively biased because of feature overlap, meaning that the distance between the two overlapping features was systematically underestimated. For the least-squares fitting to sums of model functions, however, the bias stayed within \SI{0.1}{px}.

This negative bias for center-of-mass centroiding has been described before \cite{Baumgartl2006, Ramirez-Saito2006} and is a logical consequence of the method: if two features overlap, each of the features obtains extra intensity on the inside of the dimer. This bias increases in magnitude with decreasing particle separation, until both features are detected precisely in between the two actual positions. The bias increases also with increasing mask radius $R$, as shown in Figure \ref{fig:refine_distance}.

Apart from this negative bias, we observed a longer ranged positive bias. This effect has its origin in the preprocessing. For center-of-mass centroiding, it is vital that the constant image background is subtracted. This is conventionally achieved by subtracting a rolling average of the image with box size of typically $D_{bg} = 2R + 1$ \cite{Crocker1996}. Although this method has proven to be robust for background subtraction, it also introduces a skew in the feature signals when features are closer than $\ell + D_{bg}$ (see Figure~S1). Here, $\ell$ is the typical feature diameter. From this we conclude that it is important not to use a rolling average background subtraction in order to accurately track features that are spaced closer than $\ell + D_{bg}$. If the background subtraction was omitted, the positive bias was indeed not observed, as can be seen in Figure \ref{fig:refine_distance}. In order to account for the background signal in the least-squares fitting algorithm, we introduced a background variable $B$ in the objective function (Eq. \ref{eq:global_obj}) instead.

\begin{figure*}[h]
	\centering
	\includegraphics[width=\textwidth]{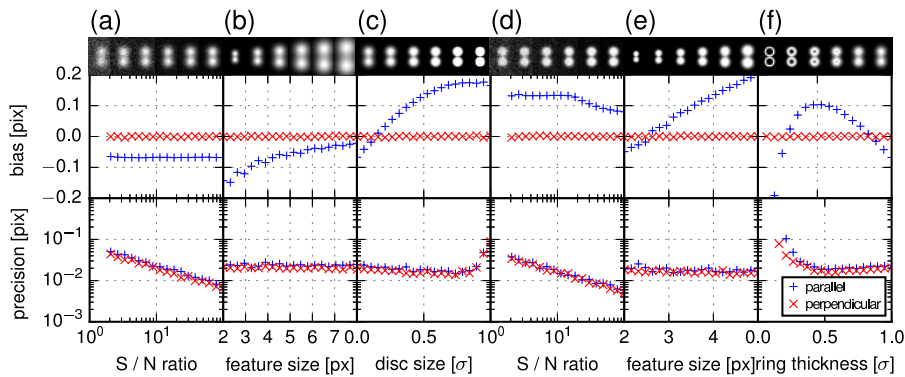}
	\caption{Tracking errors of artificial overlapping features at a separation distance of \SI{8}{px}. The top row presents the computer generated model features, the middle and bottom row show the mean deviation (bias) and the root of the central variance of the deviations (precision), respectively. The data is separated into the error parallel (blue straight crosses) and perpendicular (red tilted crosses) to the line connecting the true feature positions. The effect of (a) the signal-to-noise ratio (S/N ratio) and (b) feature radius $\sigma$ is plotted for Gaussian shaped model features. In (c) the tracking errors of overlapping solid discs with relative disc size $d$ are shown (see Eq. \ref{eq:disc}). For disc shaped features with $d = 0.5$, the effect of S/N ratio and feature size is plotted in (d) and (e), respectively. Finally in (f), the effect of relative ring thickness $t$ in a ring shaped feature is shown (see Eq. \ref{eq:ring}). Unless stated otherwise, a feature size $\sigma = \SI{4}{px}$ and signal-to-noise ratio $\mathrm{S/N} = 10$ were employed.}
	\label{fig:dimer_study}
\end{figure*}

Secondly, we analyzed the bias and precision of overlapping Gaussian features, disc shaped features, and ring shaped features while keeping the particle separation constant at \SI{8}{px}. See Figure \ref{fig:dimer_study}. In all cases, we observed no bias in the perpendicular coordinate, as is expected from the symmetry of the dimer. Also, the precision for the perpendicular direction was in close agreement with the precision of parallel direction.

In Figure \ref{fig:dimer_study}(a), it can be seen that the signal-to-noise (S/N) ratio did not influence the bias for Gaussian shaped model features, while the precision improved with increasing S/N ratio. At $\mathrm{S/N} \geq 2.0$, the least-squares optimizer was always able to find a minimum. At $\mathrm{S/N} < 2.0$, the optimizer sometimes diverged and yielded random results. This failure of least-squares fitting was reported already for $\mathrm{S/N} < 4$ by Cheezum \textit{et al.} \cite{Cheezum2001}. As the SLSQP minimization allows for bounds on the feature parameters, we were able to suppress the diverging solutions by limiting the displacements of center coordinates to the mask size $R$. This enhancement enables us to also use the least-squares method for $2 \leq \mathrm{S/N} < 4$.

In Figure \ref{fig:dimer_study}(b), it is visible that the bias in the parallel coordinate decreased with increasing feature size. Although the bias was so small that we can still speak of ``subpixel accuracy'', the bias of approximately \SI{-0.1}{px} for typical values of $\sigma$ might be problematic for super resolution techniques in which sum of Gaussians are used as model functions for overlapping point spread functions \cite{Gordon2004, Qu2004a}. As the magnitude of the bias increased with decreasing feature size and not with increasing S/N, we conclude that the bias is caused by the discretization of the feature shape, which depends on the used discretization model.

As colloidal molecules are often larger than the diffraction limit, their feature shape is typically not Gaussian. Here we will assess the effect of non-Gaussian shapes on the tracking bias and precision using a disc-shaped model feature as described by Eq. \ref{eq:disc}. See Figures \ref{fig:dimer_study}(c)-(e). The observed precision in the refined position of the overlapping discs was surprisingly high, and the precision even slightly increases up to a disc size of $0.8\sigma$ (Figure \ref{fig:dimer_study}(c)). This was probably caused by the larger integrated signal intensity of the disc shaped feature, which increased the S/N ratio integrated over the feature. For disc sizes greater than $0.8\sigma$, the precision degraded due to the absence of smooth feature edges. The bias was lowest for small feature sizes (Figure \ref{fig:dimer_study}(e)), since the disc-sized feature is then almost equal to the Gaussian shaped feature. Still, the magnitude of the bias did not exceed \SI{0.2}{px} for all tested disc-shaped features.

Finally, in Figure \ref{fig:dimer_study}(f), we tested the least-squares fitting of Gaussians on ring-shaped model features (Eq. \ref{eq:ring}), such as may be obtained for particles with fluorescent markers on their surface only. Although a Gaussian is clearly a poor model function for these ring-shaped model features, it still performed remarkably well with absolute bias and precision both below \SI{0.1}{px} for any ring thickness above $0.2\sigma$, probably because the tails of the features are still Gaussian-shaped. For thin rings with a thickness below $0.2\sigma$, the least-squares optimization diverges. For these feature shapes, a more appropriate model function should be used.

To summarize, we observed that least-squares fitting to sums of Gaussians is able to accurately refine the location of overlapping Gaussian-shaped features. The negative bias of multiple pixels present in center-of-mass centroiding is reduced to less than \SI{0.1}{px} if the feature radius is above \SI{3}{px}.
This makes fitting to sums of Gaussian an appropriate method for refining overlapping features with typical radii around \SI{4}{px} and S/N ratios above 2. Although the Gaussian is not a perfect model for disc-shaped or ring-shaped features, the bias and precision were very similar due to the limited pixel size for typical images of overlapping colloidal particles, given that the feature edges are smooth.

For overlapping features that are not well modeled by a Gaussian and that have a radius larger than \SI{5}{px}, different model functions should be used. As described by Jenkins \textit{et al.} \cite{Jenkins2008}, it is possible to experimentally obtain an average feature shape and successfully use this for feature refinement of single features. For an extension to multiple overlapping features, we found that this technique is computationally too demanding, as there are no efficient optimizers for functions with discretized parameters. In order to use this technique for overlapping features, the average feature shape should be described with a continuous function, which can be directly used in our framework for least-squares minimization.

Although an accuracy of \SI{0.1}{px} is sufficient for many applications, a further improvement in accuracy could be reached by maximizing the log-likelihood corresponding to Eq. \ref{eq:global_obj} instead of using the direct least-squares minimization. For single features, using a maximum likelihood estimator has been proven to give a more precise estimate of the true feature positions \cite{Smith2010, Abraham2009}.

\subsection{Constrained least-squares}
If additional information about the tracked features is available, constraints can be applied to increase tracking accuracy. In our framework for least-squares optimization of summed radial model functions, any combinations of parameters in the image model function (Eq. \ref{eq:generic_model}) can be constrained by equations of the following form:

\begin{equation}
g(P_n) = 0 \qquad \mathrm{or} \qquad  g(P_n) \geq 0.
\end{equation}

Here, $g$ is a function and $P_n$ is an array consisting of all parameters of features that are in a cluster of size $n$.
We demonstrate the use of constraints here using colloidal dimers with known distance between the two constituent spheres. Using our algorithm we automatically tracked 1006 out of 1170 recorded frames. A constraint was chosen such that the distance between the constituent spheres equals the average distance measured on SEM images (\SI{1.58}{\um}). The resulting tracked three-dimensional images can be seen in Supporting Video~S1.

\begin{figure}[h]
\centering
\includegraphics{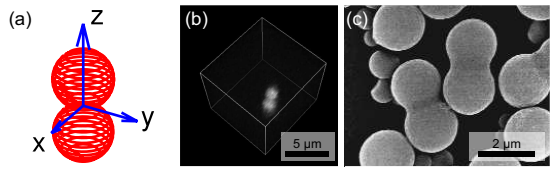}
\caption{Images of colloidal dimers. The coordinate system corresponding to the diffusion tensor is shown in (a). The origin of the coordinate system lays in the point of highest symmetry. A typical three-dimensional confocal image that is used for the particle tracking is displayed in (b), next to a representative Scanning Electron Micrograph of the employed colloidal dimers (c).}
\label{fig:clusters}
\end{figure}

As the shape of a colloidal cluster is anisotropic, the short-term diffusion of such a particle is also anisotropic: for example, a dimer has a lower hydrodynamic friction when moving along its $z$-axis, compared to when moving along its $x$-axis. In general, the dynamics of any Brownian object is described by a symmetric second-rank tensor of diffusion coefficients, consisting of 21 independent elements \cite{Kraft2013a}. 
We chose the point of highest symmetry for the origin of the cluster based coordinate system and aligned the z-axis with the long axis of the dimer, so that all off-diagonal terms in the diffusion tensor are zero.
See Figure \ref{fig:clusters}(a).
The computed diffusion tensors were averaged over lagtimes up to \SI{0.6}{\second}. The resulting diffusion tensor reflects the symmetry of the dimer and can be seen in supporting Table~S1.

\begin{table}
	\caption{Anisotropic diffusion coefficients of the colloidal dimer. The coordinate system is defined in Figure \ref{fig:clusters}. The error denotes the \SI{95}{\percent} confidence interval estimated using a bootstrap algorithm.}
		\begin{tabular}{llll}
			\hline
			Type&Axis&Diffusion coefficient\\
			\hline
			Translation&x, y& \SI{61.2+-3.9e-3}{\um\squared\per\second}\\
			Translation&z& \SI{65.2+-4.2e-3}{\um\squared\per\second}\\
			Rotation&x, y&\SI{13.0+-1.1e-3}{\per\second} \\
			\hline
		\end{tabular}
	\label{table:diffusion_coeffs}
\end{table}

In Table \ref{table:diffusion_coeffs} we summarized the measured translational and rotational diffusion coefficients of the colloidal dimer. In line with previous results from holographic microscopy measurements \cite{Fung2013}, we observed that the translational diffusion constant along z is higher than the translational coefficient along x and y. These results illustrate that our new tracking algorithm is able to compute quantitative information from microscopy images of colloidal clusters, without the need of manual corrections.

\section{Conclusion}
We have presented a new algorithm for single-particle tracking that enables automated tracking of overlapping features with high accuracy and precision. The algorithm is based on a the well-known algorithm developed by Crocker and Grier \cite{Crocker1996} and implements two improvements. First, by exploiting the information obtained from the linking already in the segmentation stage, we were able to use the history of the feature positions to obtain segmentation with significantly fewer mistakes. In a test on two-dimensional experimental data of dimers, all frames were segmented correctly, while the conventional algorithm correctly segments only one third of the frames. 

The second improvement enables sub-pixel accuracy. The conventional center-of-mass refinement is unable to find unbiased feature locations: signal overlap results in a negative bias if the feature separation distance is below the mask diameter, and the commonly used rolling average background subtraction imposes a positive bias already at separation distances below approximately 1.5 times the mask diameter. We reach sub-pixel accuracy and precision by least-squares fitting to sums of Gaussians. First, we tested Gaussian-shaped model features with varying signal-to-noise ratio and feature size and found that the obtained coordinates are biased less than \SI{0.1}{px} for a feature radius above \SI{3}{px}. The bias decreases with increasing feature size, implying that the pixel discretization is the cause.

Non-Gaussian features can also be tracked with surprising accuracy and precision using a Gaussian model function: for features with a radius of \SI{4}{px}, sub-pixel accuracy and precision was obtained even if \SI{80}{\percent} of the feature is a solid disc. For ring shaped features, sub-pixel accuracy and precision was achieved for a ring thickness of more than \SI{20}{\percent} of the feature size. For feature radii larger than \SI{5}{px}, the bias increases above \SI{0.2}{px}, implying that more appropriate model functions should be used in order to obtain sub-pixel accuracy of overlapping features.

We demonstrated the use of constraints in least squares fitting with experimental three-dimensional image sequences of colloidal dimers. Trajectories through \SI{86}{\percent} of all frames were obtained without any manual refinement. From this, the diffusion tensor was reported and found to accurately reflect the particle symmetry.

With the described method, two problems are solved that are encountered when employing conventional tracking methods to overlapping features. Firstly, the need for case-to-case meticulous optimization or manual reparation of tracks is significantly reduced. Secondly, by employing least squares fitting to summed Gaussians we found that the bias of the center-to-center separation distance is \SI{0.2}{px} in the worst case, which clearly outperforms the center-of-mass centroiding. Our method provides accurate automated tracking of videos containing overlapping features with minimal need for manual adjustments.

\section*{Acknowledgement}
The authors would like to thank Giovanni Biondaro and Vera Meester for supplying the PMMA clusters for the three-dimensional diffusion measurements. This work was supported by the Netherlands Organisation for Scientific Research (NWO/OCW), as part of the Frontiers of Nanoscience program, and through VENI grant 680-47-431.  

\section*{References}
\bibliographystyle{unsrt}

\begin{thebibliography}{10}
	
	\bibitem{Murray1996}
	C.~A. Murray and David~G. Grier.
	\newblock {Video microscopy of monodisperse colloidal systems}.
	\newblock {\em Annu. Rev. Phys. Chem.}, 47(1):421--462, 1996.
	
	\bibitem{Weeks2000}
	E.~R. Weeks, J.~C. Crocker, A.~C. Levitt, A.~Schofield, and D.~A. Weitz.
	\newblock {Three-dimensional direct imaging of structural relaxation near the
		colloidal glass transition}.
	\newblock {\em Science}, 287(5453):627--631, 2000.
	
	\bibitem{Kegel2000}
	W.~K. Kegel and A.~van Blaaderen.
	\newblock {Direct observation of dynamical heterogeneities in colloidal
		hard-sphere suspensions}.
	\newblock {\em Science}, 287:290--293, 2000.
	
	\bibitem{Dinsmore2001}
	A.~D. Dinsmore, E.~R. Weeks, V.~Prasad, A.~C. Levitt, and D.~A. Weitz.
	\newblock {Three-dimensional confocal microscopy of colloids}.
	\newblock {\em App. Optics}, 40(24):4152--9, 2001.
	
	\bibitem{Meng2014}
	G.~Meng, J.~Paulose, D.~R. Nelson, and V.~N Manoharan.
	\newblock {Elastic instability of a crystal growing on a curved surface}.
	\newblock {\em Science}, 343(6171):634--7, feb 2014.
	
	\bibitem{MacKintosh1999}
	F.~C. MacKintosh and C.~F. Schmidt.
	\newblock {Microrheology}.
	\newblock {\em Curr. Opin. Colloid In.}, 4(4):300--307, 1999.
	
	\bibitem{Tseng2002}
	Y.~Tseng, T.~P. Kole, and D.~Wirtz.
	\newblock {Micromechanical mapping of live cells by multiple-particle-tracking
		microrheology}.
	\newblock {\em Biophys. J.}, 83(6):3162--3176, 2002.
	
	\bibitem{Neuman2008}
	K.~C. Neuman and A.~Nagy.
	\newblock {Single-molecule force spectroscopy: optical tweezers, magnetic
		tweezers and atomic force microscopy}.
	\newblock {\em Nat. methods}, 5(6):491--505, 2008.
	
	\bibitem{Gelles1988}
	J.~Gelles, B.~J. Schnapp, and M.~P. Sheetz.
	\newblock {Tracking kinesin-driven movements with nanometre-scale precision}.
	\newblock {\em Nature}, 331(6155):450--453, 1988.
	
	\bibitem{Yildiz2003}
	A.~Yildiz, J.~N. Forkey, S.~A. McKinney, T.~Ha, Y.~E. Goldman, and P.~R.
	Selvin.
	\newblock {Myosin V walks hand-over-hand: single fluorophore imaging with
		1.5-nm localization}.
	\newblock {\em Science}, 300(5628):2061--2065, 2003.
	
	\bibitem{Betzig2006}
	E.~Betzig, G.~H. Patterson, R.~Sougrat, O.~W. Lindwasser, S.~Olenych, J.~S.
	Bonifacino, M.~W. Davidson, J.~Lippincott-Schwartz, and H.~F. Hess.
	\newblock {Imaging intracellular fluorescent proteins at nanometer resolution}.
	\newblock {\em Science}, 313(2006):1642--1645, 2006.
	
	\bibitem{Huang2007}
	B.~Huang, W.~Wang, M.~Bates, and X.~Zuang.
	\newblock {Three-dimensional super-resolution imaging by stochastic optical
		reconstruction microscopy}.
	\newblock {\em Science}, 319:810--813, 2007.
	
	\bibitem{Crocker1996}
	J.~C. Crocker and D.~G. Grier.
	\newblock {Methods of Digital Video Microscopy for Colloidal Studies}.
	\newblock {\em J. Colloid Interf. Sci.}, 179(1):298--310, 1996.
	
	\bibitem{Savin2005}
	T.~Savin and P.~S. Doyle.
	\newblock {Static and dynamic errors in particle tracking microrheology}.
	\newblock {\em Biophys. J.}, 88(1):623--38, jan 2005.
	
	\bibitem{Ghosh1994}
	R.~N. Ghosh and W.~W. Webb.
	\newblock {Automated detection and tracking of individual and clustered cell
		surface low density lipoprotein receptor molecules}.
	\newblock {\em Biophys. J.}, 66(5):1301--1318, 1994.
	
	\bibitem{Saxton1997a}
	M.~J. Saxton and K.~Jacobson.
	\newblock {Single-particle tracking: applications to membrane dynamics}.
	\newblock {\em Annu. Rev. Bioph. Biom.}, 26:373--399, 1997.
	
	\bibitem{Ober2004}
	R.~J. Ober, S.~Ram, and E.~S. Ward.
	\newblock {Localization accuracy in single-molecule microscopy}.
	\newblock {\em Biophys. J.}, 86(2):1185--1200, 2004.
	
	\bibitem{Smith2010}
	C.~S. Smith, N.~Joseph, B.~Rieger, and K.~A. Lidke.
	\newblock {Fast, single-molecule localization that achieves theoretically
		minimum uncertainty}.
	\newblock {\em Nat. Methods}, 7(5):373--5, 2010.
	
	\bibitem{Manoharan2003}
	V.~N. Manoharan, M.~T. Elsesser, and D.~J. Pine.
	\newblock {Dense packing and symmetry in small clusters of microspheres}.
	\newblock {\em Science}, 3010:483--487, 2003.
	
	\bibitem{Kraft2009}
	D.~J. Kraft, J.~Groenewold, and W.~K. Kegel.
	\newblock {Colloidal molecules with well-controlled bond angles}.
	\newblock {\em Soft Matter}, 5(20):3823, 2009.
	
	\bibitem{Meester2016}
	V.~Meester, R.~W. Verweij, C.~van~der Wel, and D.~J. Kraft.
	\newblock {Colloidal recycling: reconfiguration of random aggregates into
		patchy particles}.
	\newblock {\em ACS Nano}, 10:4322--4329, 2016.
	
	\bibitem{Glotzer2007}
	S.~C. Glotzer and M.~J. Solomon.
	\newblock {Anisotropy of building blocks and their assembly into complex
		structures}.
	\newblock {\em Nat. Mater.}, 6(7):557--562, 2007.
	
	\bibitem{Hunter2011}
	G.~L. Hunter, K.~V. Edmond, M.~T. Elsesser, and E.~R. Weeks.
	\newblock {Tracking Rotational Diffusion of Colloidal Clusters}.
	\newblock {\em Opt. Commun.}, 19(18):17189--17202, 2011.
	
	\bibitem{Edmond2012a}
	K.~V. Edmond, M.~T. Elsesser, G.~L. Hunter, D.~J. Pine, and E.~R. Weeks.
	\newblock {Decoupling of rotational and translational diffusion in supercooled
		colloidal fluids}.
	\newblock {\em P. Natl. Acad. Sci. USA}, 109(44):17891--17896, 2012.
	
	\bibitem{Cheezum2001}
	M.~K. Cheezum, W.~F. Walker, and W.~H. Guilford.
	\newblock {Quantitative comparison of algorithms for tracking single
		fluorescent particles}.
	\newblock {\em Biophys. J.}, 81(4):2378--2388, 2001.
	
	\bibitem{Jenkins2008}
	M.~C. Jenkins and S.~U. Egelhaaf.
	\newblock {Confocal microscopy of colloidal particles: towards reliable,
		optimum coordinates}.
	\newblock {\em Adv. Colloid Interfac.}, 136:65--92, 2008.
	
	\bibitem{VanBlaaderen1995}
	A.~van Blaaderen and P.~Wiltzius.
	\newblock {Real-space structure of colloidal hard-sphere glasses}.
	\newblock {\em Science}, 270(19):1177--1179, 1995.
	
	\bibitem{Besseling2009}
	R.~Besseling, L.~Isa, E.~R. Weeks, and W.~C.~K. Poon.
	\newblock {Quantitative imaging of colloidal flows}.
	\newblock {\em Adv. Colloid Interfac.}, 146:1--17, 2009.
	
	\bibitem{Serge2008}
	A.~Serg{\'{e}}, N.~Bertaux, H.~Rigneault, and D.~Marguet.
	\newblock {Dynamic multiple-target tracing to probe spatiotemporal cartography
		of cell membranes}.
	\newblock {\em Nat. Methods}, 5(8):687--694, 2008.
	
	\bibitem{Jaqaman2008}
	K.~Jaqaman, D.~Loerke, M.~Mettlen, H.~Kuwata, S.~Grinstein, S.~L. Schmid, and
	G.~Danuser.
	\newblock {Robust single-particle tracking in live-cell time-lapse sequences}.
	\newblock {\em Nat. Methods}, 5(8):695--702, 2008.
	
	\bibitem{Baumgartl2006}
	J.~B{\"{a}}umgartl, J.~L. Arauz-Lara, and C.~Bechinger.
	\newblock {Like charge attraction in confinement: Myth or truth?}
	\newblock {\em Soft Matter}, 2(8):631, 2006.
	
	\bibitem{Gordon2004}
	M.~P. Gordon, T.~Ha, and P.~R. Selvin.
	\newblock {Single-molecule high-resolution imaging with photobleaching}.
	\newblock {\em P. Natl. Acad. Sci. USA}, 101(17):6462--6465, 2004.
	
	\bibitem{Qu2004a}
	X.~Qu, D.~Wu, L.~Mets, and N.~F. Scherer.
	\newblock {Nanometer-localized multiple single-molecule fluorescence
		microscopy}.
	\newblock {\em P. Natl. Acad. Sci. USA}, 101(31):11298--303, 2004.
	
	\bibitem{trackpy031}
	D.~Allan, T.~Caswell, N.~Keim, and C.~van~der Wel.
	\newblock {Trackpy v0.3.1}.
	\newblock {\em Zenodo}, 55143, 2016.
	
	\bibitem{Maneewongvatana1999}
	S.~Maneewongvatana and D.~M. Mount.
	\newblock {It's okay to be skinny, if your friends are fat}.
	\newblock {\em Proceedings of the 4th Annual Workshop on Computational
		Geometry}, pages 1--8, 1999.
	
	\bibitem{Kraft1988}
	D.~Kraft.
	\newblock {A software package for sequential quadratic programming}.
	\newblock Technical report, DLR German Aerospace Center, DFVLR-FB 88-28, 1988.
	
	\bibitem{scipy}
	E.~Jones, T.~Oliphant, P.~Peterson, and Others.
	\newblock {SciPy: Open source scientific tools for Python}.
	
	\bibitem{Besseling2015}
	T.~H. Besseling, J.~Jose, and A.~van Blaaderen.
	\newblock {Methods to calibrate and scale axial distances in confocal
		microscopy as a function of refractive index}.
	\newblock {\em J. Microsc.}, 257(2):142--150, 2015.
	
	\bibitem{VanderWel2016a}
	C.~van~der Wel, R.~K.~Bhan, R.~W. Verweij, H.~C. Frijters, A.~D. Hollingsworth,
	S.~Sacanna, and D.~J. Kraft.
	\newblock {Preparation of colloidal organosilica spheres through spontaneous
		emulsification}, {\em in preparation}.
	
	\bibitem{Kraft2013a}
	D.~J. Kraft, R.~Wittkowski, B.~{Ten Hagen}, K.~V. Edmond, D.~J. Pine, and
	H.~L{\"{o}}wen.
	\newblock {Brownian motion and the hydrodynamic friction tensor for colloidal
		particles of complex shape}.
	\newblock {\em Phys. Rev. E}, 88(5):2--6, 2013.
	
	\bibitem{Ramirez-Saito2006}
	A.~Ram{\'{i}}rez-Saito, C.~Bechinger, and J.~L. Arauz-Lara.
	\newblock {Optical microscopy measurement of pair correlation functions}.
	\newblock {\em Phys. Rev. E}, 74(3):030401, 2006.
	
	\bibitem{Abraham2009}
	A.~V. Abraham, S.~Ram, J.~Chao, E.~S. Ward, and R.~J. Ober.
	\newblock {Quantitative study of single molecule location estimation
		techniques.}
	\newblock {\em Opt. Express}, 17(26):23352--73, 2009.
	
	\bibitem{Fung2013}
	J.~Fung and V.~N. Manoharan.
	\newblock {Holographic measurements of anisotropic three-dimensional diffusion
		of colloidal clusters}.
	\newblock {\em Phys. Rev. E}, 88:020302, 2013.
	
\end{thebibliography}

\newpage

\section*{Supporting Material for ``Automated tracking of colloidal clusters with sub-pixel accuracy and precision''}

\large Authors: Casper van der Wel and Daniela J. Kraft
\setcounter{figure}{0}
\setcounter{table}{0}
\makeatletter
\renewcommand{\fnum@figure}{Figure S\thefigure}
\renewcommand{\fnum@table}{Table S\thetable}
\renewcommand{\theequation}{S\arabic{equation}}
\makeatother

\begin{figure}[h]
	\centering
	\includegraphics{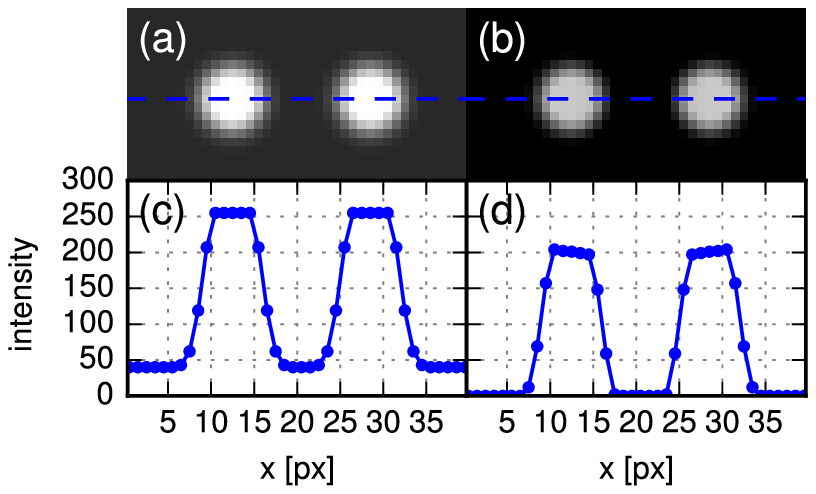}
	\caption{Illustration of the positive bias due to background subtraction. If the image background is nonzero (a), it can be subtracted using a rolling average resulting in a perfectly black background (b). However, in the image cross sections (c) and (d), it can be seen that the rolling average also results in a skew of the feature shapes, which gives a outwards directed bias.}
	\label{fig:boxcar_bias}
\end{figure}

\begin{table}[h]
	\centering
	\caption{Tensor of dimer diffusion coefficients. The translational coefficients are given in units of \SI{e-3}{\um\squared\per\second}, the rotational coefficients in units of \SI{e-3}{\per\second}, and the rotation-translation cross terms in units of \SI{e-3}{\um\per\second}. Because rotation around the z-axis cannot be measured for a dimer, we omitted the corresponding elements. The error denotes the \SI{95}{\percent} confidence interval estimated using a bootstrap algorithm.}
	\label{table:dimer_tensor}
	\begin{tabular}{l|rrrrr}
		&x&y&z&$\theta_x$&$\theta_y$\\\hline
		x&61.6$\pm$4.0&-0.9$\pm$2.8&-0.4$\pm$3.1&0.0$\pm$1.3&-0.4$\pm$1.3\\
		y&-0.9$\pm$2.8&60.8$\pm$3.8&-0.7$\pm$3.0&-0.4$\pm$1.3&-0.2$\pm$1.4\\
		z&-0.4$\pm$3.1&-0.7$\pm$3.0&65.2$\pm$4.2&-0.0$\pm$1.3&-0.4$\pm$1.4\\
		$\theta_x$&0.0$\pm$1.3&-0.4$\pm$1.3&-0.0$\pm$1.3&12.5$\pm$1.1&-0.2$\pm$0.7\\
		$\theta_y$&-0.4$\pm$1.3&-0.2$\pm$1.4&-0.4$\pm$1.4&-0.2$\pm$0.7&13.4$\pm$1.1\\
	\end{tabular}
\end{table}

\setcounter{figure}{0}
\makeatletter
\renewcommand{\fnum@figure}{Still of Video S\thefigure}
\makeatother

\begin{figure}[h]
	\centering
	\includegraphics{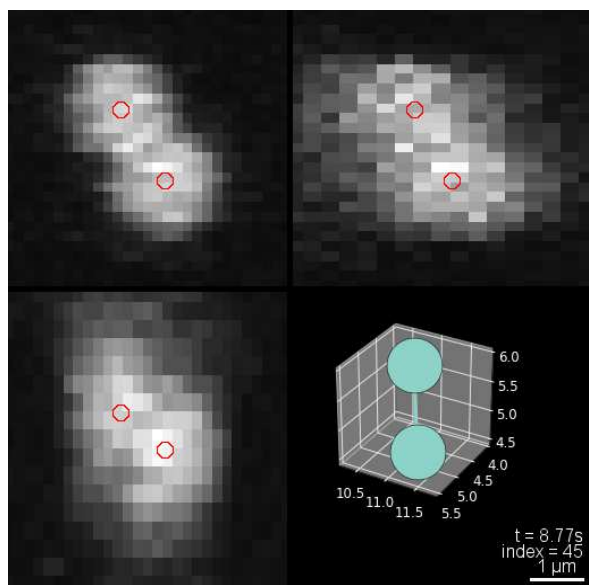}
	\caption{Three-dimensional video of a diffusing colloidal dimer as measured by confocal microscopy. The particle tracking is shown in an overlay on maximum intensity projections in three directions (upper left: xy, upper right: xz, lower left: yz). The rectangular pixels reflect the larger pixel size in the z direction. On the lower right, a three-dimensional plot is showing the orientation of the dimer. The axes are in microns.}
	\label{fig:video_still}
\end{figure}

\end{document}